\documentclass{article}
\usepackage[utf8]{inputenc}
\usepackage[a4paper, total={6.5in, 10in}]{geometry}
\usepackage{hyperref}
\usepackage{graphicx}
\graphicspath{ {images/} }
\usepackage{amsmath}
\usepackage{amssymb}
\usepackage{color, colortbl}
\usepackage[dvipsnames]{xcolor}
\hypersetup{colorlinks,breaklinks,urlcolor=purple, linkcolor=cyan, citecolor = blue}
\usepackage{enumitem}
\usepackage{comment}
\usepackage{mathrsfs}
\usepackage{bbm}
\usepackage{titlesec}
\usepackage[ruled,vlined]{algorithm2e}
\usepackage{authblk}
\usepackage{setspace}
\usepackage{multirow}
\allowdisplaybreaks
\title{\textbf{Finding Optimal Cancer Treatment using Markov Decision Process to Improve Overall Health and Quality of Life}}
\author[1]{\textbf{Navonil Deb}}
\author[2]{\textbf{Abhinandan Dalal}}
\affil[1,2]{Master of Statistics, Final year, Indian Statistical Institute, Kolkata, India}
\author[3]{\textbf{Gopal Krishna Basak}}

\affil[3]{Theoretical Statistics and Mathematics Unit, Indian Statistical Institute, Kolkata, India}
\date{}
\begin{document}

\maketitle


\begin{abstract}
    Markov Decision Processes and Dynamic Treatment Regimes have grown increasingly popular in the treatment of diseases, including cancer. However, cancer treatment often impacts quality of life drastically, and people often fail to take treatments that are sustainable, affordable and can be adhered to. In this paper, we emphasize the usage of ambient factors like profession, radioactive exposure, food habits on the treatment choice, keeping in mind that the aim is not just to relieve the patient of his disease, but rather to maximize his overall physical, social and mental well being. We delineate a general framework which can directly incorporate a net benefit function from a physician as well as patient's utility, and can incorporate the varying probabilities of exposure and survival of patients of varying medical profiles. We also show by simulations that the optimal choice of actions often is sensitive to extraneous factors, like the financial status of a person (as a proxy for the affordability of treatment), and that these actions should be welcome keeping in mind the overall quality of life.
\end{abstract}



\section{Introduction}

Cancer is a major public health problem worldwide and is also the second leading cause of morbidity and mortality in the United States. For nearly 1 in every 7 deaths around the world are accounted by cancer, claiming more lives than AIDS, tuberculosis and Malaria combined. In most of the developing lower and middle-income countries, people still do not have access to a well regulated and organized system for cancer diagnosis. With the adoption of urbanized lifestyle and changing food habits in the economically transitioning nations, the risk of death due to cancer is expected to increase as it has already been predicted that by 2030, the global threat of cancer will grow to 20.7 million new cancer cases, with the death of 13 million cancer patients because of growing and ageing population and the developing nations will have a major part in that \cite{siegel2016}. The International Agency for Research on Cancer GLOBOCAN project has predicted that India’s cancer burden will nearly double in the next 20 years, from slightly over a million new cases in 2012 to more than 1.7  million by 2035 \cite{mallath2014growing}.

Diagnosis of cancer is currently one of the most discussed topics in the realms of medical and biological research. Existing treatment modalities for cancer include surgery, chemotherapy, radiotherapy, immunotherapy, medication or surveillance among others. Even if a cancer patient is generally treated with only two or three of these treatment modalities, receiving them in different order and time points might result in different outcomes. Unfortunately decisions concerned to multimodal treatment is often dependent on the patient’s clinical experience and therefore answering questions like- what is the optimal sequence of applying these modalities or what should be the optimal time gap among them, often becomes  difficult for the medical practitioners. Perhaps such challenges may be overcome with suitably designed methodologies that suggest with the treatment policy, a sequence of treatment related decisions, which optimizes the overall health-economic welfare of the patient. One possible approach of finding such an optimal treatment policy is to search for the optimal treatment path from the set of all possible paths by brute-force but finding such policies can be computationally expensive and often impractical. Even an intuitive or heuristic search for the optimal treatment modality may sometimes be computationally inefficient \cite{schaefer2005modeling}. Earlier approaches in this direction were due to Beil and Wein (2001) \cite{beil2001analysis} and Hathout et al. (2016) \cite{hathout2016modeling} by using ordinary and partial differential equations respectively. Their models were able to give success in cases where medical practitioners usually restrict themselves to surgical treatments along with radiotherapeutic procedures. 

These challenges represent significant opportunities for improvement. In medical science, several decision theoretic approaches have already been tried out for diagnosis of various diseases as well as cancer. Bennett and Hauser (2013) used an artificial intelligent simulation based framework for clinical decision making of cancer therapies \cite{bennett2013artificial} and later Maass and Kim \cite{maass2020markov} proposed a Markov decision process based approach for treating cancer with multi-modal therapies .

The goal of this paper is to develop a clinical decision making model which along with an individual patient’s medical records also considers her/his personalized attributes and aims to maximize a suitably proposed measure of ‘reward’. In this article, a patient’s ‘reward’ or ‘payoff’, in a heuristic sense, shall be used to incorporate the idea of her/his life expectancy, along with her/his ability, following the treatment, to return to the lifestyle s/he had before being diagnosed with the disease (from decision theoretic aspect, a reward can be thought of as a function having the reverse nature of a loss function). Here, we propose a rationally chosen real-valued reward function. For the sake of computational simplicity and tractability, our model assumes that the nature of our model is ‘Markovian’ i.e. memoryless (one-step dependence where future given the present situation is independent of the past). Section \ref{qol} shall discuss the importance of treatment fine tuned to everyday life of a patient. In section \ref{model}, we formulate the model and explicitly state our model assumptions. In section \ref{estimation}, we focus on the estimation of parameters in our model with the help of data on covariates. Section \ref{simulation} will focus on a simulation based study of this model and will try to interpret the results and draw meaningful inferences from them. Section \ref{conclusion} will be the conclusion and remarks of this paper.


\section{Treatment fine-tuned to everyday life}\label{qol}

We would like to emphasize the importance of everyday life factors in cancer treatment. As we leap the gap of making personalized medicine from an elusive dream to an imminent reality \cite{lesko2007personalized}, we must not restrict ourselves to only pharmacogenomic and pharmacogenetic information \cite{jain2002personalized}, but also other exogenous factors, like a lifestyle, race, food habits etc. For instance, Mu{\~n}oz et. al \cite{munoz2001case} finds human diet to be a significant factor to be affecting risk of gastric cancer in a case-control study in Venezuela.  In fact, he reports a potential rise in chances of gastric cancer with increased consumption of cereals, starchy vegetables, dairy products and fruits. Even when controlled for creatinine levels, sodium levels were found to have a positive impact on gastric cancer risks, which implies that an increase in salt intake turns out to be a strong risk factor. The study also finds relations between alcohol consumption and gastric cancer. 

Smoking and tobacco habits have a strong impact on the lung cancer chances of an individual, as illustrated in many studies, for example by Loeb et. al. \cite{loeb1984smoking}.  However, lung cancer risks go beyond mere active smoking habits. As Correa et. al illustrate \cite{correa1983passive}, non-smokers married to heavy smoker spouses have an elevated risk of lung cancer. The study also posits questions about maternal smoking habits and impacts they have on the children. 

Risks of cancer are not only limited to personal habits, but also environmental factors. Increased exposure to pollution levels, proximity to urban areas and industrial hubs pose a threat of lung cancer to the inhabitants. Nafstad et. al \cite{nafstad2003lung} found such evidence in Norwegian men, while Nyberg et. al \cite{nyberg2000urban} found similar results in Stockholm. Factors like SO$_2$ concentration, SPM levels affect lung cancer risks heavily. From Jedrychowski et. al.’s study \cite{jedrychowski1990case} in Poland to Tseng et. al’s study in Taiwan \cite{tseng2019relationship}, the ubiquity of these effects indeed show a dramatic impact on lung cancer risks. In fact, even after diagnosis, these ambient air pollution levels shorten survival chances and survival lengths of patients, as found in a study by Eckel et. al. \cite{eckel2016air}


In fact, the risk effects of cancer are not only limited to everyday activities, but some effects have been found on racial effects and socio-economic status too. Early studies by the likes of Bradley et. al. \cite{bradley2002race}, Schwartz et. al. \cite{schwartz2009interplay} and Le et. al \cite{le2008effects} support such findings.  Particularly, African-American individuals tend to have a significant deviation from the mean rate of incidence across races. The effects have been present in case-control studies, in region specific  as well as general studies, also across wide range of cancer types, including but not limited to, gastric, colorectal, lung and breast. Mu{\~no}z et. al. \cite{munoz2001case} in fact studies the education levels of the subjects as well, and finds that more educated people are at a lesser risk of exposure. All these are related to the consciousness of people about the disease and ways to reduce their risk of incidence. However, although later studies debate the racial effects, as by Yang et. al. \cite{yang2010racial}, the effects of socio-economic status on the incidence and survival rates have been well pronounced in most studies. In fact, factors like profession of a subject can also affect the exposure risks of subjects. For example, mine workers have been found to have greater incidence to lung cancer, as by Peller \cite{peller1939lung}, and Navarro et. al \cite{fernandez2012proximity}. In fact, this is not confined to a particular mine, but rather among a wide spectrum of mine workers, as haematite \cite{chen1990mortality}, taconite \cite{allen2015cancer}, gold \cite{obiri2006cancer}, arsenic, nickel and chromium \cite{witschi2001short}. The effect is not limited to the current mining workers only. Residual effects have been found by Woodward et. al. \cite{woodward1991radon} in former mine workers of Radium Hill uranium mine due to exposure to radon daughters. Mortality rates have been studied in dust exposed mine and pottery workers as well \cite{chen1992mortality}. 

The effects of socio-economic status and demographic variables have not been limited to just exposures, but chances of survival as well. Affordability of expensive treatment has been a great concern in the global arena \cite{aggarwal2014affordability}, and there have been a plethora of attempts to bridge the gap between the treatment and the treated. Instances of such attempts include sustainable pricing of drugs \cite{uyl2018sustainability}, health insurances \cite{zhao2020health} and affirmative action \cite{leech2020balancing}. Goldstein et. al. \cite{goldstein2017global} studies the cost affordability of drugs on the global platform, and finds that cancer treatment is most expensive in the United States, while in India, it is unaffordable to the general masses by a large margin. Moye et. al \cite{moye2020availability} studies this scenario in middle income countries like Mexico, and the broad takeaway is that cancer drugs continue to remain greatly unaffordable in middle income countries, bringing down the global survival rate. With growing wealth inequality in the world, health inequality has also been on the rise, and there have been various attempts to quantify and measure this inequality \cite{allison2004measuring, wagstaff2003difference}. Sarah and Curtis \cite{curtis1998there} also study the place of geography in health inequality, and this shall also have an impact on the medicare a person receives in treatment.  


The overall health of a patient is not only characterized by the absence of a disease and infirmity, but rather as a state of physical, mental and social well-being, as regarded by the World Health Organization \cite{world1995constitution}. Thus, alongside factors affecting cancer risks and survival rates, we also need to consider the quality of life parameters in the treatment of an individual. Testa and Simonson \cite{testa1996assessment} has a fantastic discussion on the importance of quality of life factors, and ways to assess and measure it. For our purpose, consider a working individual, aged around mid-40s, who has been diagnosed with cancer. If s/he has adequate affordability with respect to the ambient financial situation, then the physician should concentrate on speedy recovery of her/his health. The treatment may be expensive, but if it rewarded with her/him being able to return to work after recovery, s/he has the potential to revive the expenses and live a fulfilling life. Thus, the treatment profile should be chosen optimally keeping in mind the person’s ability to return to work, and regaining his lifestyle prior to being diagnosed with cancer. On the other hand, consider an aged individual post-retirement. Under such a scenario, it is likely that heavy treatments and the subsequent side-effects have adverse (and often fatal impacts) on her/his health. Even if the treatment is not fatal and relieves her/him of cancer, the potential pitfalls lead her/him to being bed-ridden for the rest of his life. The physician then was successful of relieving her/him of cancer, but the treatment profile performed very poorly on the quality of life scale. Such scales must be used hand in hand with the disease relieving intent of the physicians, so as to make healthcare much more affordable and fulfilling. Quality of life parameters in medicare and survival rates of cancer have been growing popular in recent times, for example in \cite{brock1993quality}, and we welcome this development and emphasize its usage in reward based dynamic treatment regimes.

Dropouts have been a rampant phenomenon in several medical treatments, and cancer is no exception. Factors like depression, lack of understanding physician’s instructors, and income and financial barriers have contributed to the dropout of patients from cancer treatments. In fact, people often do not realize they have dropped out of treatment due to their inability to comprehend the concept of adherence to treatment, as found by Wells et. al. \cite{wells2011cancer}. Ethnic minorities and financially challenged people are often at the forefront of the challenge of adherence to treatment. Another aspect of people who are prone to dropouts are those who are too ill to travel, as suggested in \cite{applebaum2012factors}, and telemedicine approaches would be beneficial to them. 

Quality of life approaches have been growing popular in several diseases, for instance, McRae et. al. \cite{mcrae2004effects} looks into this index for Parkinson’s disease patients. An important aspect that has risen with this QOL approach is the growth of behavioural therapy. For example, Monnikes et. al. \cite{ heymann2000combination} demonstrate the superiority of behavioural therapy and medical treatment over medical treatment alone, and it is time we try out such behavioural aspects in cancer treatment as well. As Testa and Simonson \cite{testa1996assessment} elaborate, such QOL based treatment needs to ensure safety, efficacy and convenience on the treatment aspect, and compliance on the behavioural aspect of the patient. The outcome parameters should include risk reduction, years of healthy life rewarded, overall quality of life, summed together in the form of a net benefit function. The MDP approach we shall be outlining in this paper, provides a readymade position to use this net-benefit function, henceforth referred to as the reward function, and a framework to stochastically optimize it through optimal choice of actions. In fact, by using this approach to demonstrate the impact on the overall well-being of life of the patients, new simple yet effective nudges can be developed, as often argued by Richard Thaler \cite{thaler2009nudge}. With growing concerns on health inequality and consequent policy implications \cite{deaton2002policy}, exploring such aspects would grow indispensible in the near future.

\section{Model Formulation: Modeling Cancer as a Markov Decision Process}\label{model}

In our literature, we implicitly assume that the transition between different stages (malignancy) is a Markovian process i.e. the future transition is only dependent on the current stage of the tumor, but not on the past information. Several authors like \cite{kim2009markov, imani2020markov} have made such assumptions to suggest probabilistic decision models pertinent to this problem. Therefore, we aim to formulate a probabilistic decision model with the assumption of Markovian transitions. While defining the notation and the problem we adopt the conventions of \cite{puterman}.

A \textbf{Markov process} is a stochastic process with the \textit{memoryless} property, that is, given information of present  the future realizations of the random process does not depend upon the past realizations. If $\{ X_n \}_{n=0}^\infty$ is the random process, then it is a Markov process if for $\mathbb{P}(X_{n-1}=x_{n-1},\cdots, X_0=x_0) > 0$
\begin{equation}\label{markov}
    \mathbb{P}(X_n = x_{n}\mid X_{n-1}=x_{n-1},\cdots, X_0=x_0) = \mathbb{P}(X_n=x_n\mid X_{n-1}=x_{n-1})
\end{equation}

In a nutshell, A \textbf{Markov Decision Process} (MDP) is a Markov process where certain amount of reward is given (or cost deducted) after an action is performed. In general, the process might be any history dependent stochastic process but our model will stick to the Markovian assumption for computational ease.

\subsection{The Problem at Hand}
Consider a subject under a course of treatment for a period $T$ and s/he seeks for an optimal treatment decision. Here \textbf{optimal} decision denotes to the one which maximizes the expected value of reward. Following are the different components of the model:
\subsubsection{State Space (Stage of Cancer)}

For a particular type of cancer, we divide the type of malignancies by $\{S_1, \cdots, S_n\}$ and consider this set as the state space $S$. If we consider more than one type of cancer then each of the states can be represented by $S_i^{(j)}$ where $i=1,\cdots, n;\ \text{and}\ j=1\cdots,K$ (number of cancer types). 

\subsubsection{Action Space (Possible Treatments)}

The action space is the set of actions, may depend on time or may not, to be performed when the system is at a particular state at a given time. To reduce computational complexity, in this model we would consider the action space to be time homogeneous i.e. when a patient is at a certain state the treatment modalities would not depend on the time point of the observation, it will only depend on the corresponding stat. Therefore when the system, at time t, is at the state s, the set of actions is $\mathcal{A}_{s,t}=\mathcal{A}_s$. The action space is given by: $\mathcal{A}=\cup_{s\in S}\mathcal{A}_s$

\subsubsection{Transition Probabilities}

When the system is at state $i$ and after taking action $a$ at time $t$ it makes a transition to the state $j$, the transition probability is denoted by:
\begin{equation}\label{transprob}
    p_t(j\mid i,a) = \mathbb{P}(X_t = j\mid X_{t-1}=i, a_t=a)
\end{equation}

The transition probability is a time dependent function i.e. under same action and same transition, probabilities may be different at different time points. For simplicity time homogeneity is assumed in many cases.

\subsubsection{Reward Function}
Each transition offers a reward (or deducts some cost which can be thought of as a negative reward) depending upon the state of the system and the corresponding action taken from that state. A reward function is denoted by $r_t(a,i,j)$ where the transition occurs at time $t$ under action $a$ from a state $i$ to state $j$. The term \textit{reward} is referred to as some sort of measure for patient's ability to lead a normal lifestyle after s/he is treated against cancer. While discussing a simulation based example in further section, we will interpret the results for a class of reward functions and argue in support of that. The reward function is considered as the \textbf{gain} when the system is in state $s$ at time $t$ by taking expectation over the probability distribution over the transitioning state, that is
\begin{equation}\label{expreward}
    r_t(s, a) = \mathbb{E}\left[ r_t(a,s,s') \right]=\sum_{s'\in S}r_t(a,s,s')\hspace{0.1cm} p_t(s'\mid s,a)
\end{equation}

We would consider the reward function to be dependent on certain parameters which represent the socio-economic or geographic background of the corresponding subject. For persons belonging to different geographic regions and beholding different socio-economic status may have different reward functions even though other variables controlling the rewards are fixed. For example, \cite{zajacova2015employment} illustrates how cancer diagnosis has substantial effects on the economic well-being of affected adults and their families, whereas according to \cite{pisani1992breast}, two different persons belonging to widely separated geographic regions may not face the same risk of cancer due to the genetic factor of the population or the climatic factor.

\subsubsection{Decision Rule and Policy}
In decision theory, a non-randomized decision rule is a function from the sample space to action space i.e. for a decision rule $d$, sample space $\mathcal{X}$ and action space $\mathcal{A}$, $d:\mathcal{X}\mapsto \mathcal{A}$. A randomized decision rule $\delta$ is a probability distribution over the set of non-randomized decision rules $\mathcal{D}$ that is $\delta\in \mathcal{P(D)}$. 

In reference to the model we propose, a decision rule is a function from the state space to action space i.e. a decision rule takes the state of the system as an input and decides which action to take. We confine our model to class of deterministic decision rules i.e. at time $t$, the decision rule $d_t: S \mapsto \mathcal{A}$. The \textit{Markovian} nature of our model ensures that the decision at time $t$, depends only on the current time and state but does not account for the previous chain of states and action i.e. the rules are independent of the history. A \textbf{policy} $\pi$ is defined as the collection of decision rules over time points i.e. $\pi = (d_1,d_2,\cdots,d_{T-1})$.

\subsection{Optimization Goal}

In this model, the utility of reward functions will be considered to be additive in nature. Hence, we take $\psi(r_1, \cdots, r_T) = \sum_{i=1}^T r_i$

Let $\Pi$ be the space of all policies. The main objective is to determine a policy $\pi^*=(d_1^*,\cdots,d_{T-1}^*)$ such that the expected utility of rewards is maximized (Assume such a policy exists). Under a policy $\pi$ and initial state $s$, the expected total utility upto the horizon $T$ is:
\begin{equation}\label{exputil}
    v_T^{\pi}(s)=\mathbb{E}_s^{\pi}\left[\sum_{t=1}^{T-1}r_t(s_t,d_t(s_t))+r_T(s_T)\right]
\end{equation}

We seek a policy $\pi^*$ such that the expected utility of the cancer patient is maximized such that
\[ v_T^{\pi^*}(s) \geq v_T^{\pi}(s)\quad \forall s\in S, \pi \in \Pi\]
In order to search for such a policy, we apply \textbf{Backward Induction Algorithm} on the set of reward equations obtained in each step of the algorithm, also known as \textbf{Bellman's Optimality Equations}.

\begin{algorithm}[H]\label{policyeval}
\SetAlgoLined
\KwResult{Optimal Policy}
set $t=N$\;
Calculate $u_N^*(s_N) = r_N(s_N)$ for all $s_N\in S$\;
\eIf{$t=1$}{
        \textbf{stop};
    }
    {
        set $t\leftarrow t-1$\;
         $\forall s_t\in S$ calculate\qquad  $u_t(s_t, a) = r_t(s_t,a) + \sum_{j\in S} p_t(j\mid s_t, a)u_{t+1}^*(j)\ \text{for all}\ a\in \mathcal{A}_{s_t}$
         $$ u_t^*(s_t) =  \max_{a \in \mathcal{A}_{s_t}} u_t(s_t, a), \quad \mathcal{A}_{s_t,t}^* = \arg \max_{a \in \mathcal{A}_{s_t}} u_t(s_t, a) $$
         \textbf{continue}\;
    }
    \caption{Backward Induction Algorithm (Markovian Setup)}
\end{algorithm}

\section{Model Formulation: Statistical Model for Estimation of Parameters}\label{estimation}

The model and methods we have so far introduced does not take into account the features of the patient in hand. Though some of the factors might get included by virtue of influencing the reward function (for example, the weight of an individual is not directly taken into account, but might contribute to the overall final health of the patient, which might be of interest for the final reward function); there might be some factors which, though not directly influencing the reward function, can act as significant contributors in making a decision. So we introduce those variables, who do not influence the reward function, but can influence the outcomes of the transitions. 

Thus, in this section, we introduce the idea of \textbf{covariates}, and intend to extend the model taking into account the covariate realizations possessed by the patient. In general, covariates are independent variables which contribute in the underlying model as observations/measurements and can be treated as control variables. We present two ideas to use them in the procedure.

\subsection{Non-adaptive Approach: Fixed Covariates and Unchanged State Space}
Given a state $s_t$ and action $a_t$ at time $t$, we have used the probabilities $p_t(s|s_t,a_t)$ for all $s\in S$ by direct estimation from data. However, not taking in the covariate profile of the patients makes the estimated transition probability for a randomly chosen patient from the patient population at state $s_t$ at time $t$, on which $a_t$ action is applied. 
\par Now instead, we use a \textbf{Proportional Odds Cumulative Logit Model}, popularized by McCulagh \cite{mccullagh}. The rationale behind using this model is that the states, in the way they are defined, have an inherent order of severity. Let there be $p$ covariates $X_1,\cdots,X_p$. These covariates are assumed to be fixed quantities and differ for each individual. But for a fixed patient the value of the covariate is given i.e., the data of the covariates are not random quantities.
\par Suppose we are at time $t$. If we have $n$ individuals i.e. $n$ patients who are at state $s_t$ and action $a_t$ is applied on them, the data is given by $\{X_{ij}:1\leq i\leq p, 1\leq j\leq n\}$. The data matrix (or design matrix) is given by $\mathbf{X}=\left(\left(X_{ij}\right)\right)_{1\leq i\leq p, \newline 1\leq j\leq n}$. Let $|S|=J$, i.e. we have $J$ many categories for the state variable and they are ordinal in nature. Our goal is to find out how many items belong to the $k$\textsuperscript{th} category or below where $1\leq k\leq J$.

The model is:

\begin{equation}\label{propcumlogit}
    \log\left[\frac{\pi_1^{(t,s_t,a_t)} + \cdots + \pi_j^{(t,s_t,a_t)}}{\pi_{j+1}^{(t,s_t,a_t)} + \cdots + \pi_J^{(t,s_t,a_t)}}\right] =  \alpha_j^{(t,s_t,a_t)} + \beta_1^{(t,s_t,a_t)} x_1 + \cdots + \beta_p^{(t,s_t,a_t)} x_p
\end{equation}

where, $j\in \{1, \cdots, J-1\}, \pi_j^{(t,s_t,a_t)} = \mathbb{P}(s_{t+1} = j \mid s_t, a_t, \mathbf{x}), \mathbf{x} = \left( x_1,\cdots, x_p \right)^\top$
\vskip 0.2cm
The model can alternatively written as:
\begin{equation}
    \text{logit}\left( \mathbb{P}(s_{t+1} \leq j \mid s_t, a_t, \mathbf{x}) \right) = \alpha_j^{(t,s_t,a_t)} + \beta_1^{(t,s_t,a_t)} x_1 + \cdots + \beta_p^{(t,s_t,a_t)} x_p
\end{equation}

We shall continue our discussion assuming this model, however, the framework can be used with any model for the probabilities. Note that, the beta coefficients are not assumed to be varying with $j$. This often seems to be a restrictive assumption, which people want to do away with. Peterson and Harell \cite{partial} discuss a partial proportional model in this regard. Ari and Yildiz \cite{erkan2014parallel} has a discussion on the parallel line assumption of the model, while Ananth and Kleinbaum \cite{ananth1997regression} provides a comprehensive review of several models for ordinal response variable as used here. 

The estimation of parameters can be done using maximum likelihood estimation with the help of \textbf{Fisher's scoring method}. It is to be noted that instead of proportional odds cumulative logit model, the multinomial logit model could have been used but there could have been two issues: (1) In multinomial logit model, for each state $j (j<J)$ and given $(t, s_t, a_t)$, there are $(p+1)$ parameters to be estimated and hence total number of parameters is $JT(J-1)(p+1)|\mathcal{A}|$ which might be large if we have any of $J, T, |\mathcal{A}|, p$ large and as a result some issues regarding \textit{curse of dimensionality} might arise. To get rid of such issues we can impose some simplistic assumtions e.g. the parameters are time homogeneous, and we use the proportional odds cumulative logit model. Therefore, the number of parameters boils down to $Jp|\mathcal{A}|+J(J-1)|\mathcal{A}| = J(p+J-1)|\mathcal{A}|$ which is less than the number of parameteres in multinomial logit model. Often, more sophisticated methods of estimation are used, for example Mukherjee et. al. \cite{mukherjee2008fitting} discusses an approach by binary collapsing of the ordinal scale, and amalgamating conditional likelihoods hence produced. However, we do not digress to discuss more sophisticated methods, as our aim here is to look into the optimal action profiles. 

\subsection{Adaptive Approach: Random Covariates and Modified State Space}

Here we allow the covariates to change their value over time. Suppose again there are $p$
covariates $X_1, \cdots, X_p$, and each can take values from a joint distribution $F$. Suppose for now that the sample space $\Omega$ of this $F$ is finite, say of order $m_1, \cdots, m_p$, and each cell has some
probability of occurrence (possibly zero). We take this F over the population of patients (and not the general demographic population). We modify the state space to be $\tilde{S} = S \times \Omega$, which
is now of the order $|S|\times |\Omega|$.

Now we allow the transition in $S$ to be from a state $i\in S$ and covariate profile $x\in\Omega$ 
to state $j\in S$ and covariate profile $y\in\Omega$, the probabilities of which can be estimated directly from the data.

The reward function for this new Markov chain is such that it remains relevant to our original reward functions. Since we are indifferent between two covariate profiles given our initial and final states, we take the stepwise reward function invariant over the covariate profiles, i.e,

\[ \forall t, \forall s, j\in S, \forall a_t\in A_s, r_t(s, a_t, j, \mathbf{x}) = r_t(s, a_t, j) \hspace{0.3cm} \forall x\in \Omega \]

The final reward is similarly taken indifferent in which covariate profile the patient finally arrives at, as long as it is the same $s\in S$.

With these state space and transition matrices in hand, we use the same procedure of backward induction and Bellman's equations, we obtain the optimal action for each $s\in S$. Note that though the reward functions remain unchanged, the probability of transition gets updated with information on the covariate profiles, and hence the optimal actions may change subject to the knowledge of covariate profile.

\subsection{Comparing the Two Approaches}

Both the approaches come with its advantages and disadvantages over the other. The first
comparison is in the philosophy of the covariates being random, the Frequentist versus the Bayesian approaches. 

However from a practical point of view, there are a few more technicalities to wonder about besides the controversy between the philosophies. The first approach allows the $\mathbf{x}$ to
be anything, i.e, it does not restrict its sample space to be finite. Hence it takes in a model assumption and gives the $\beta$'s estimates, from which probabilities can be estimated for any covariate profile $\mathbf{x}$, not just the $\mathbf{x}$'s observed.

However, the second approach, does not take in any assumption on the model for the probabilities. But it does not allow all values of covariate profile $\mathbf{x}$ in the sense that its sample space is restricted to be finite. If not, the state space becomes infinite (and possibly uncountable), which then requires closed sets and assumptions on the reward functions for the supremum to be achievable and optimal actions to be obtained using the Bellman's equations. Thus the approach loses its inherent simplicity. 

The first approach is non-adaptive in its behavior to the covariate profile in the sense that its only input is the covariate profile of the patient at the time he was first diagnosed. But the second approach allows for the covariate profile to change over the duration of treatment, a feature often desired for treatment. However, there are often covariates which naturally cannot change over time (like a genetic trait of the person) and allowing for it to change although it cannot, can make the model unnecessarily cumbersome and compromise in tractability. As a remark, we would like to comment that if exact values of the covariate profile at the beginning stage are required, like the weight of the patient at first diagnosis, then the first approach is more handy, while if information like ``overweight", ``normal weight" and ``underweight" are adequate, but, their fluctuations over time are important, one might resort to the second approach.

\section{Simulations}\label{simulation}
Acute myeloid leukaemia (AML), a type of Leukemia (blood cancer), is a clonal haemopoietic stem cell disorder in which increased proliferation and failure of differentiation in the stem cell compartment results in the accumulation of non-functional immature cells termed myeloblasts.  AML accounts for approximately 25\% of all leukaemias and is the most common form of acute leukaemia seen in adults \cite{deschler2006acute}. There has been therapeutic improvement and supportive care for AML patients for the last few decades. However, despite high complete remission rates being achieved with modern induction chemotherapy, overall survival remains at only about 50\% for those who achieve complete remission, or 26\% overall [1].  In this simulation based example, we have considered an example of 500 subjects of AML where measurements regarding every patient’s age (evidence in \cite{holmes1981cancer}), systolic blood pressure, indicator of her/his radioactive exposure (which might be determined from the information about whether that individual experienced the amount of exposure greater than a certain value in Rontgen or not), the level of some controlling hormones (e.g. Thyroxine) and presence of marker genes (e.g. CD34+ and CD38-). The information on these variables have been treated as covariates and have been used to estimate the state transition probabilities using the methods of section \ref{estimation}. The covariate structures we are simulating to train our model are following:
\begin{align*}
    & \text{Age} \sim \mathcal{N}(50,3)\\
    &\text{Blood pressure}\sim \mathcal{N}(\text{age} + 60, 0.7)\\
    & \mathbbm{1}\{\text{Radioactive exposure}\}\sim \text{Bernoulli}(0.1)\\
    & \text{Hormone level} \sim \mathcal{N}(700,20)\\
    & \text{Income} (\theta) \sim 10000 + 10^6\times \text{Pareto}(100, 10)
\end{align*}
We have considered two types of treatment modalities: 1. Observational surveillance or remission of the patient, and\ 2. Chemotherapeutic treatments. For Leukemia patients, one problem with the second type of therapy is that Chemotherapy can damage the bone marrow, which reduces the number of blood cells causing side effects. For this reason, medical practitioners always recommend using chemotherapy in cycles- a treatment period followed by a rest period. Also, we consider that the state space consists of 3 states which are labeled by 1, 2 \& 3 standing for different levels of malignancies in an increasing order of severity.

The horizon of the treatment procedure is considered to be $N = 8$ in this case. The action matrix $A$ contains $(i,j)$\textsuperscript{th} entry as the optimal treatment action to be taken by the subject if s/he is at $j$\textsuperscript{th} state at $i$\textsuperscript{th} time point. Here we consider that no action is taken when the subject is being observed at the horizon $N$ and therefore the dimension of the action matrix $A$ is $(N-1)\times S$.

For a fixed i and j, the $(i,j)$\textsuperscript{th} entry of $A$ can be modeled with a covariate $X_k$ to quantify the amount of sensitivity of $A_{ij}$ to a slight change of $X_k$. In practical cases, the covariate $X_k$ might have a certain (known) distribution and depending upon that $A_{ij}$’s sensitivity might be drastically influenced, or not affected at all. For a fixed value of $X_k$, we model $A_{ij}$ as: $\pi_{ij}^{k}(x)=\mathbbm{P}(A_{ij} = 1 \mid X_k = x) $. 

We use our stepwise reward function as follows:
$$ r_t(i,j,a) = \frac{g(i-j)}{(t+1)^2} - \frac{C_a}{\theta}\cdot e^{-\lambda t} $$

Where $C_a$ is a cost associated with action $a$ (if the action $a$ is remission, the cost is 0), $\theta$ is the income of the individual and $\lambda$ is a constant parameter considered for the decay of the reward due to an individual's to higher financial status. We set $\lambda = 1.2$ and $C_2 = 5000$. Other parameter $g$ that accounts for a  geographic factor of the reward function (for both stepwise and final) have been set to $g = 0.7$. It can be interpreted that the first term of the reward function is a result of the state transitions whereas the second term captures the effect of the action taken at time $t$ as well as the patient's financial status. The final reward function is given by (with $|S| = J$ as was defined in section \ref{estimation})
$$ r_N(j) = \frac{g(J-j)}{(N+1)^2}$$

In this case, $J=3$. Using proportional odds cumulative logit model, we estimate $\pi^k_{ij}(x)$ as $\hat\pi^k_{ij}(x)$. The usual 95\% asymptotic confidence band for the function $\pi^k_{ij}$ is given by

$$ \left[ \hat{\pi}_{ij}^{k}(x) \pm z_{n, 0.025}\sqrt{\frac{\hat{\pi}_{ij}^{k}(x)(1-\hat{\pi}_{ij}^{k}(x))}{n}} \right]$$

The covariate sensitivity plots for different covariates are provided below. For one specific covariate, say the age of a patient, we consider a range of possible values and for each value we simulate other covariates from the same training distribution, and plot the proportion of times treatment action (i.e. action 2) has been declared as the optimal action at some $(t,i)$\textsuperscript{th} entry of the action matrix ($1\leq t\leq N,\ 1\leq i\leq J$), that is treatment action is optimal at time $t$ for the patient with that particular covariate profile and suffering from the $i$\textsuperscript{th} stage of cancer. 

\begin{figure}[ht]
    \centering
    \includegraphics[width=.45\textwidth]{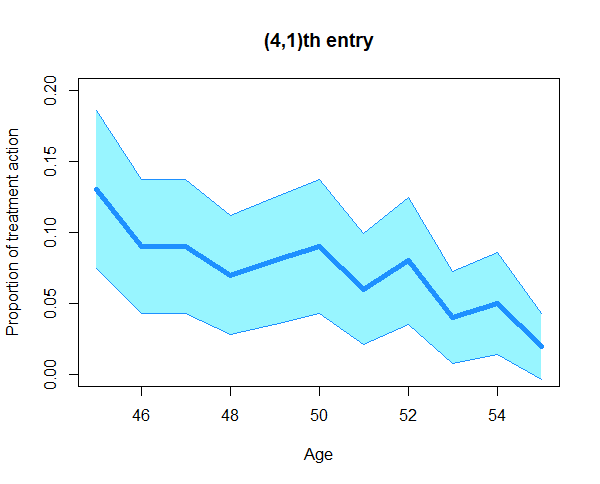}
    \centering
    \includegraphics[width=.45\textwidth]{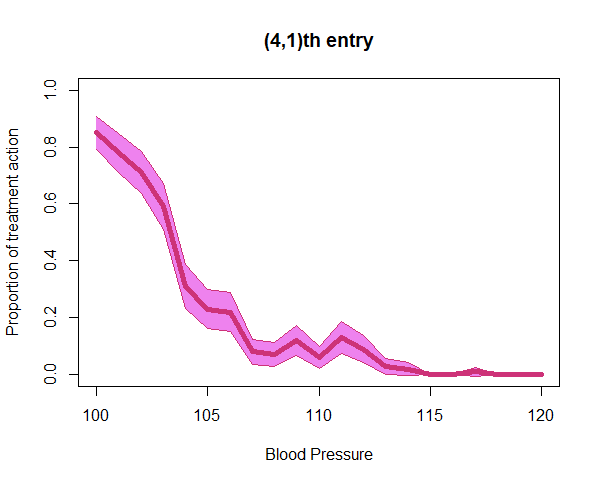}
    \caption{Sensitivity of treatment action at $A_{4,1}$ vs covariate values}\label{4,1}
\end{figure}

\begin{figure}[ht]
    \centering
    \includegraphics[width=.45\textwidth]{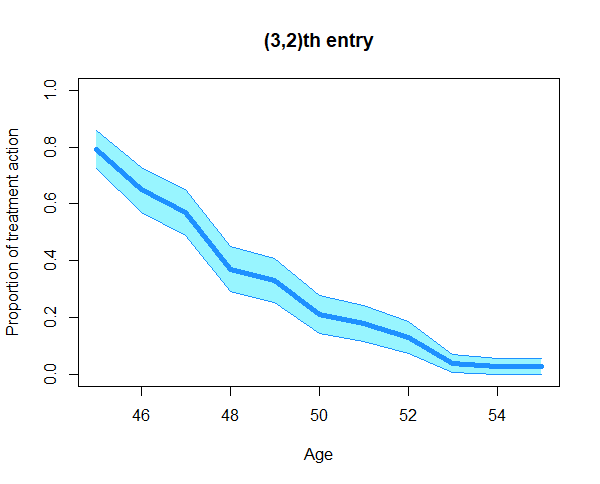}
    \centering
    \includegraphics[width=.45\textwidth]{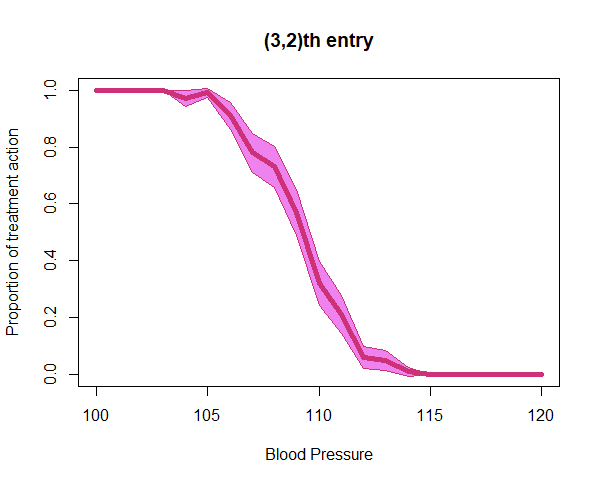}
    \caption{Sensitivity of treatment action at $A_{3,2}$ vs covariate values}\label{3,2}
\end{figure}

\begin{figure}[ht]
    \centering
    \includegraphics[width=.45\textwidth]{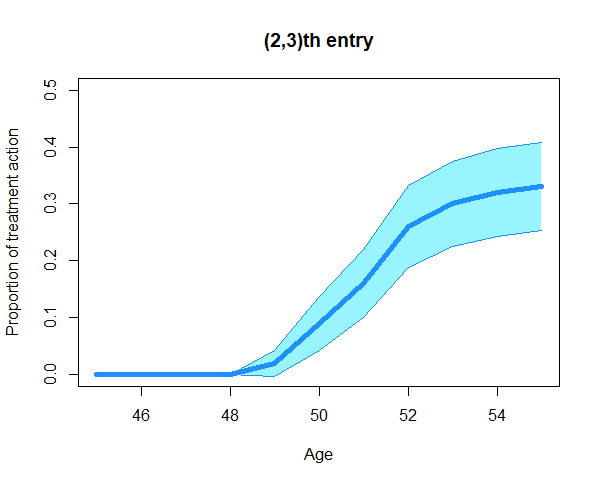}
    \centering
    \includegraphics[width=.45\textwidth]{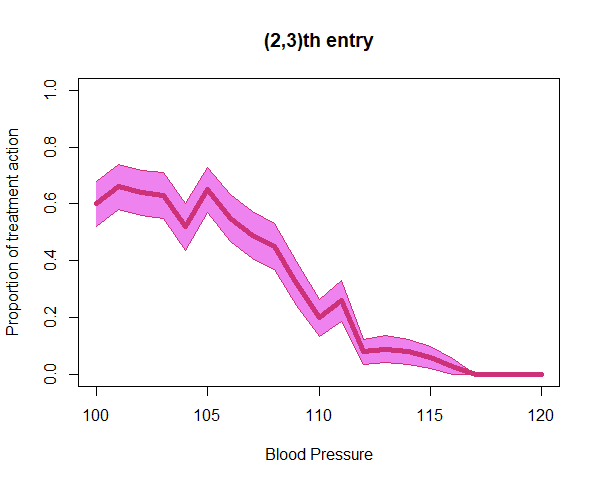}
    \caption{Sensitivity of treatment action at $A_{2,3}$ vs covariate values}\label{2,3}
\end{figure}

\begin{figure}[ht]
    \centering
    \includegraphics[width=.45\textwidth]{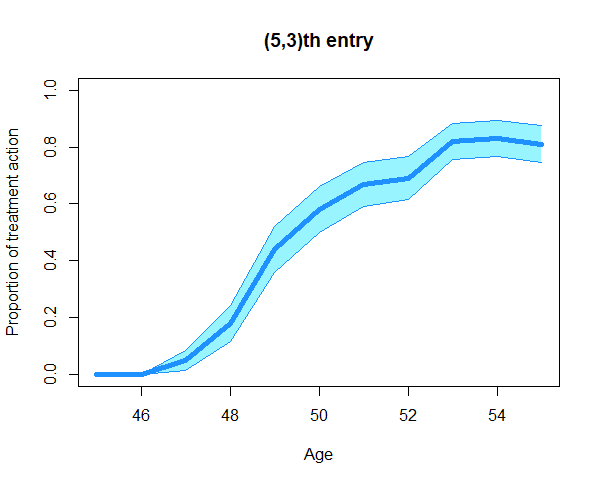}
    \centering
    \includegraphics[width=.45\textwidth]{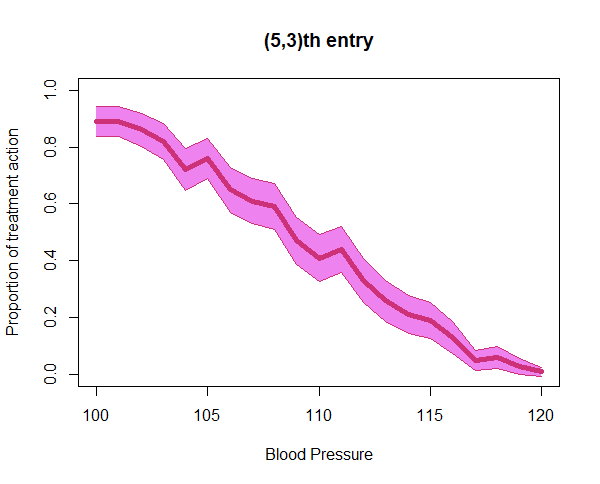}
    \caption{Sensitivity of treatment action at $A_{5,3}$ vs covariate values}\label{5,3}
\end{figure}

As suggested by figures \ref{4,1} and \ref{3,2}, when a patient is at stage 1 or stage 2 of the cancer, the optimal action is suggested as the treatment action more often if the patient is younger. But figures \ref{2,3} and \ref{5,3} indicates that if the patient is suffering from stage 3 cancer then the optimal action turns out to the treatment action more often for older patients. This type of behaviour of the action proportion suggests that for less severe stages of cancer it is not optimal to treat elder patients with powerful medications like radiotherapy, chemotherapy etc. But when the patient is suffering from stage 3, then the optimal action is indeed the treatment action which suggests some sort of retaliation against the severity of the malignancy.

For the other covariate i.e. blood pressure the sensitivity of treatment action across the range of covariate values is somewhat similar across different stages of cancer. That is, if the patient under consideration has high blood pressure then in any case the treatment action has less chance to be suggested as the optimal action by our methodology. Figures \ref{4,1}, \ref{3,2}, \ref{2,3} and \ref{5,3} capture the sensitivity of treatment action against blood pressure for some of the entries of the action matrix and all have a decreasing pattern over the range of blood pressure.

\begin{figure}[ht]
    \centering
    \includegraphics[width=0.47\textwidth]{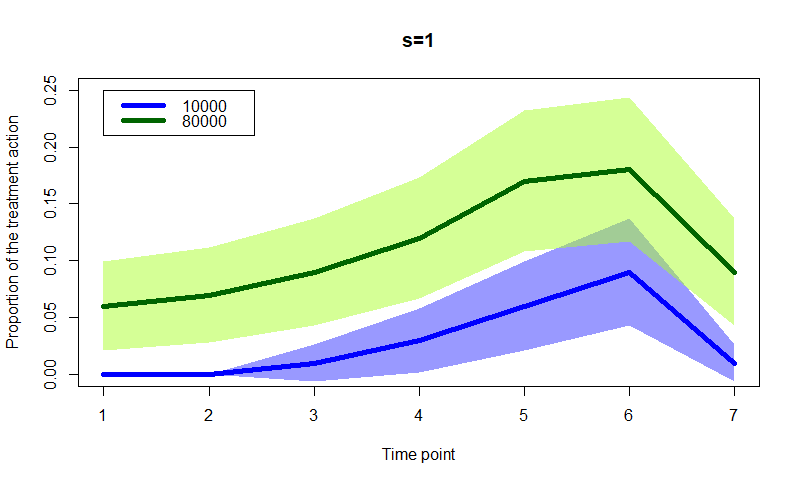}
    \includegraphics[width=0.47\textwidth]{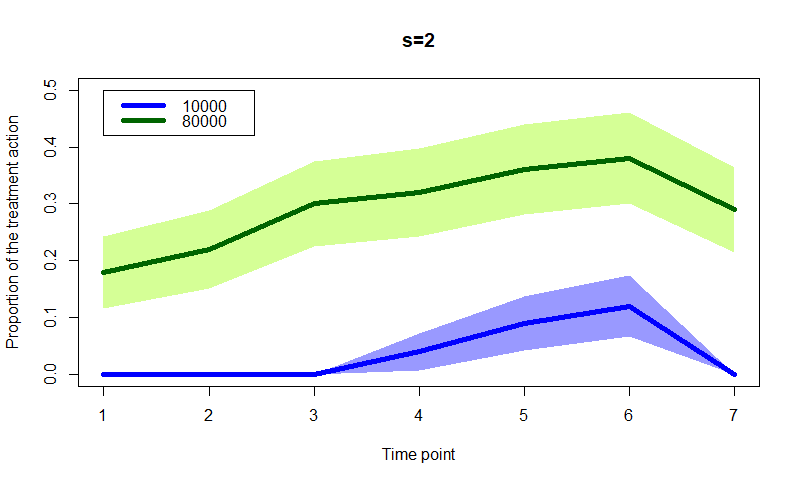}
    \includegraphics[width=0.47\textwidth]{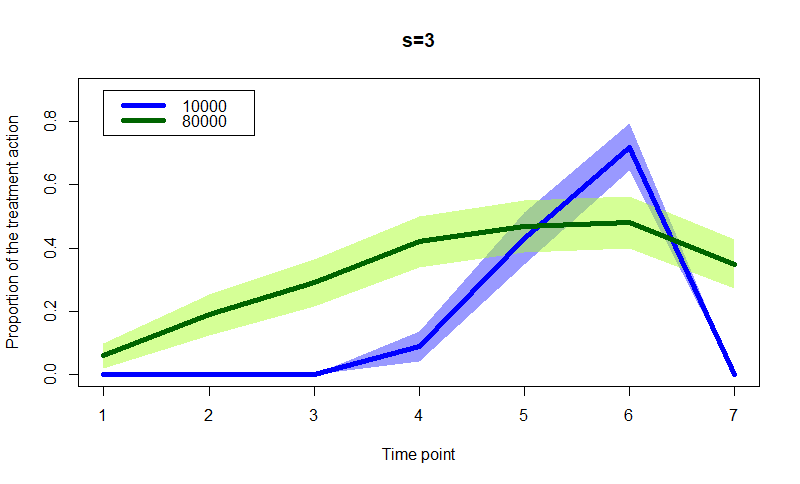}
    \caption{Proportion of treatment action vs observation time for patients belonging to income group Rs. 10000 and Rs. 80000 respectively}
    \label{10vs80}
\end{figure}

\begin{table}[ht]
    \centering
\begin{tabular}{c| c c c|c c c}
\hline\hline
\multirow{3}{*}{\begin{tabular}[c]{@{}c@{}}Time\\ ($N=8$)\end{tabular}} & \multicolumn{6}{c}{Cancer stages across income groups}                    
\\ \cline{2-7} & \multicolumn{3}{c|}{Rs. 10000} & \multicolumn{3}{c}{Rs. 80000} \\ \cline{2-7} 
 & stage 1  & stage 2  & stage 3  & stage 1  & stage 2  & stage 3  \\ \hline
1                                                                       & 0.00     & 0.00     & 0.00    & 0.06     & 0.18     & 0.06     \\ 
2                                                                       & 0.00     & 0.00     & 0.00     & 0.07     & 0.22     & 0.19     \\ 
3                                                                       & 0.01     & 0.00     & 0.00     & 0.09     & 0.30     & 0.29     \\ 
4                                                                       & 0.03     & 0.04     & 0.09     & 0.12     & 0.32     & 0.42     \\ 
5                                                                       & 0.06     & 0.09     & 0.43     & 0.17     & 0.36     & 0.47     \\ 
6                                                                       & 0.09     & 0.12     & 0.72     & 0.18     & 0.38     & 0.48     \\ 
7                                                                       & 0.01     & 0.00     & 0.00     & 0.09     & 0.29     & 0.35     \\ \hline\hline
\end{tabular}

    \caption{Proportion of action 2 (the treatment action) across different time points of treatment and stages of cancer for income groups Rs. 10000 and Rs. 80000 respectively}
    \label{table1}
\end{table}

\begin{figure}[ht]
    \centering
    \includegraphics[width=0.47\textwidth]{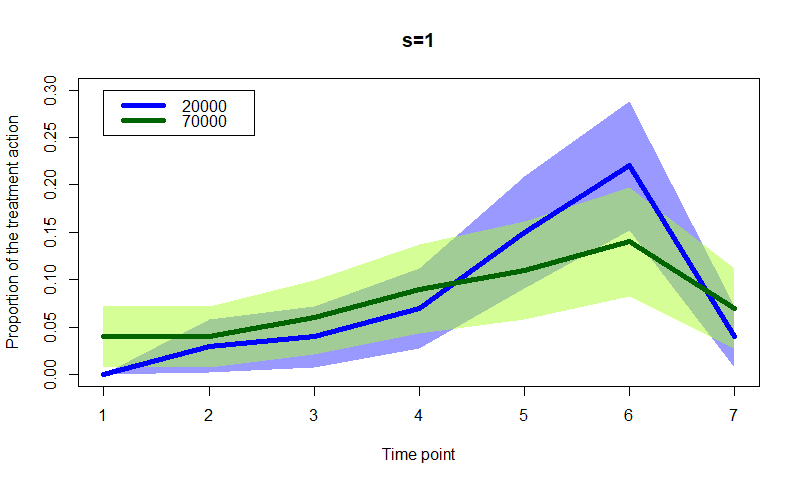}
    \includegraphics[width=0.47\textwidth]{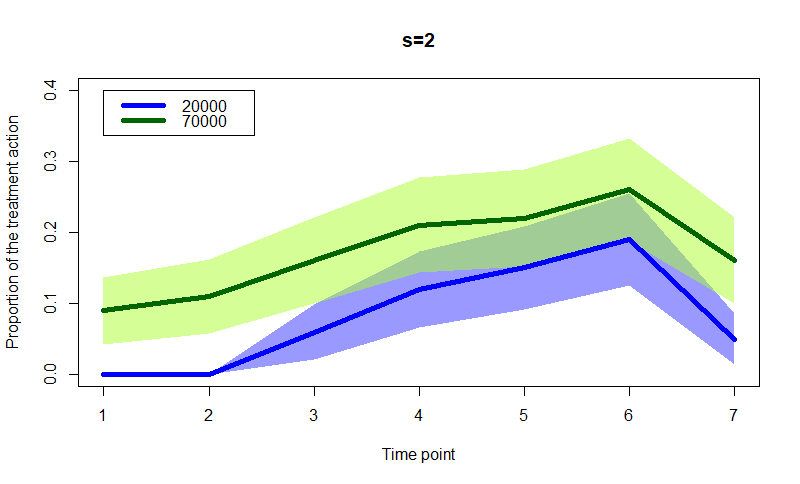}
    \includegraphics[width=0.47\textwidth]{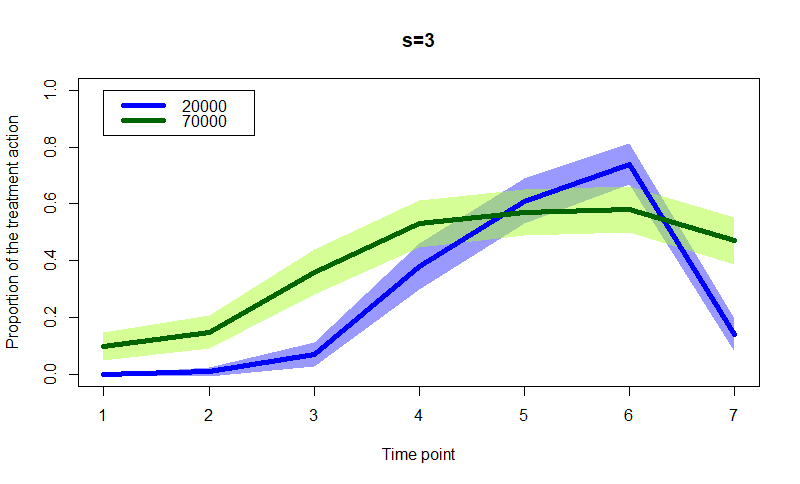}
    \caption{Proportion of treatment action vs observation time for patients belonging to income group Rs. 20000 and Rs. 70000 respectively}
    \label{20vs70}
\end{figure}

\begin{table}[ht]
    \centering
\begin{tabular}{c| c c c|c c c}
\hline\hline
\multirow{3}{*}{\begin{tabular}[c]{@{}c@{}}Time\\ ($N=8$)\end{tabular}} & \multicolumn{6}{c}{Cancer stages across income groups}                    
\\ \cline{2-7} & \multicolumn{3}{c|}{Rs. 20000} & \multicolumn{3}{c}{Rs. 70000} \\ \cline{2-7} 
 & stage 1  & stage 2  & stage 3  & stage 1  & stage 2  & stage 3  \\ \hline
1                                                                       & 0.00     & 0.00     & 0.00    & 0.04     & 0.09     & 0.10     \\ 
2                                                                       & 0.03    & 0.00     & 0.01     & 0.04     & 0.11     & 0.15     \\ 
3                                                                       & 0.04     & 0.06     & 0.07     & 0.06     & 0.16     & 0.36     \\ 
4                                                                       & 0.07     & 0.12     & 0.38     & 0.09     & 0.21     & 0.53     \\ 
5                                                                       & 0.15     & 0.15     & 0.61     & 0.11     & 0.22     & 0.57     \\ 
6                                                                       & 0.22     & 0.19     & 0.74     & 0.14     & 0.26     & 0.58     \\ 
7                                                                       & 0.04     & 0.05     & 0.14     & 0.07     & 0.16     & 0.47     \\ \hline\hline
\end{tabular}

    \caption{Proportion of action 2 (the treatment action) across different time points of treatment and stages of cancer for income groups Rs. 20000 and Rs. 70000 respectively}
    \label{table2}
\end{table}

\begin{figure}[ht]
    \centering
    \includegraphics[width=0.47\textwidth]{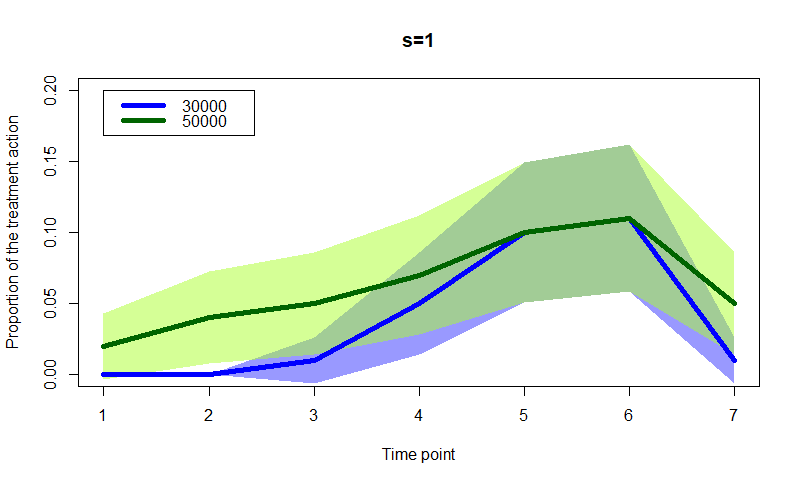}
    \includegraphics[width=0.47\textwidth]{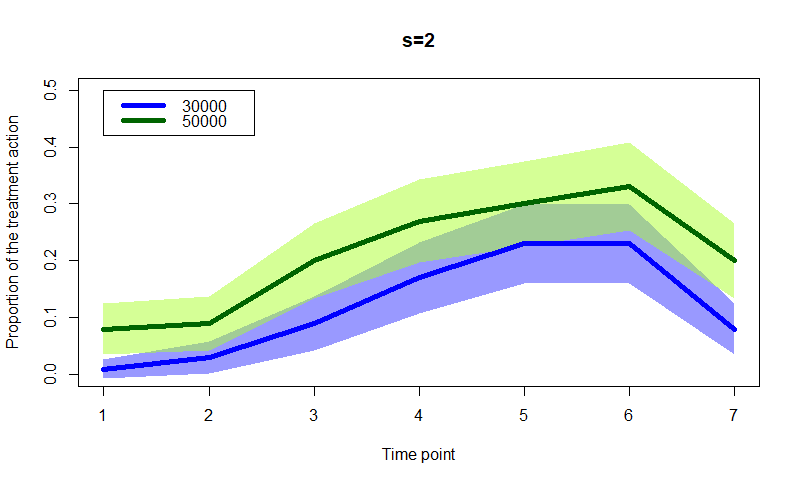}
    \includegraphics[width=0.47\textwidth]{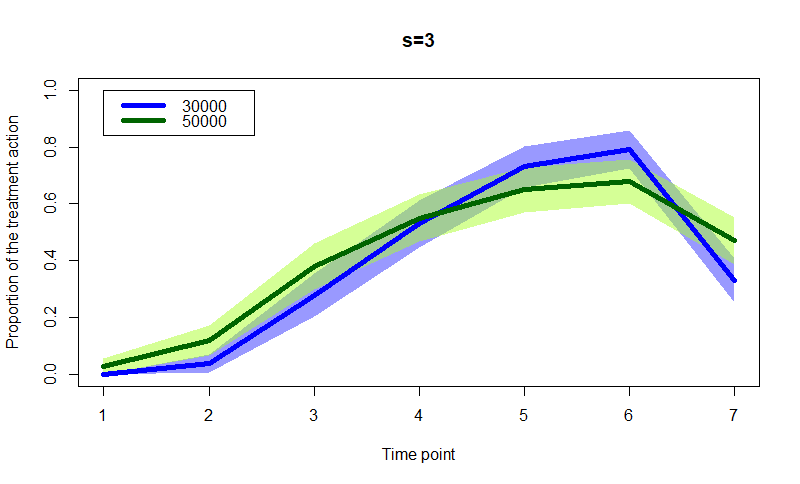}
    \caption{Proportion of treatment action vs observation time for patients belonging to income group Rs. 30000 and Rs. 50000 respectively}
    \label{30vs50}
\end{figure}

\begin{table}[ht]
    \centering
\begin{tabular}{c| c c c|c c c}
\hline\hline
\multirow{3}{*}{\begin{tabular}[c]{@{}c@{}}Time\\ ($N=8$)\end{tabular}} & \multicolumn{6}{c}{Cancer stages across income groups}                    
\\ \cline{2-7} & \multicolumn{3}{c|}{Rs. 30000} & \multicolumn{3}{c}{Rs. 50000} \\ \cline{2-7} 
 & stage 1  & stage 2  & stage 3  & stage 1  & stage 2  & stage 3  \\ \hline
1                                                                       & 0.00     & 0.01     & 0.00    & 0.02     & 0.08     & 0.03     \\ 
2                                                                       & 0.00     & 0.03     & 0.04     & 0.04     & 0.09     & 0.12     \\ 
3                                                                       & 0.01     & 0.09     & 0.28     & 0.05     & 0.20     & 0.38     \\ 
4                                                                       & 0.05     & 0.17     & 0.53     & 0.07     & 0.27     & 0.55     \\ 
5                                                                       & 0.10     & 0.23     & 0.73     & 0.10     & 0.30     & 0.65     \\ 
6                                                                       & 0.11     & 0.22     & 0.79     & 0.11     & 0.33     & 0.68     \\ 
7                                                                       & 0.01     & 0.08     & 0.33     & 0.05     & 0.20     & 0.47     \\ \hline\hline
\end{tabular}

    \caption{Proportion of action 2 (the treatment action) across different time points of treatment and stages of cancer for income groups Rs. 30000 and Rs. 50000 respectively}
    \label{table3}
\end{table}

\begin{figure}[ht]
    \centering
    \includegraphics[width=0.47\textwidth]{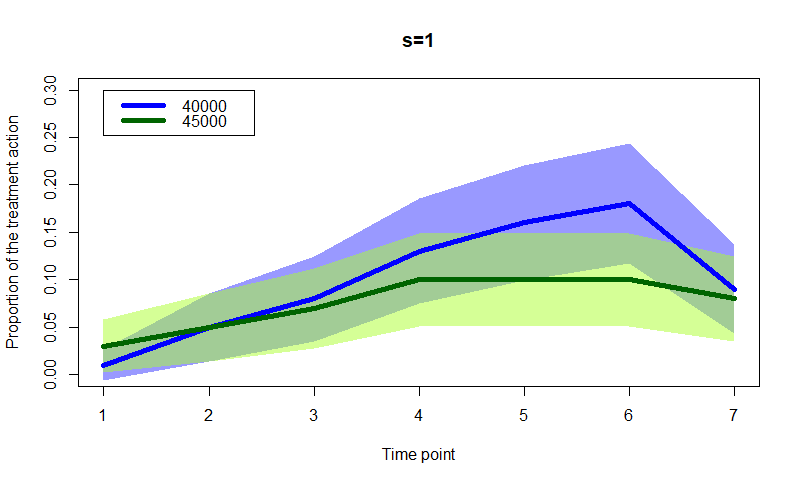}
    \label{40vs45_1.png}
    \includegraphics[width=0.47\textwidth]{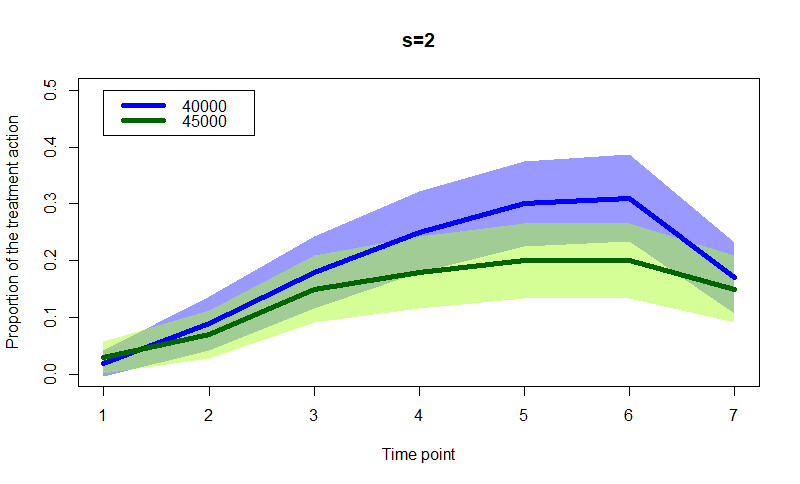}
    \includegraphics[width=0.47\textwidth]{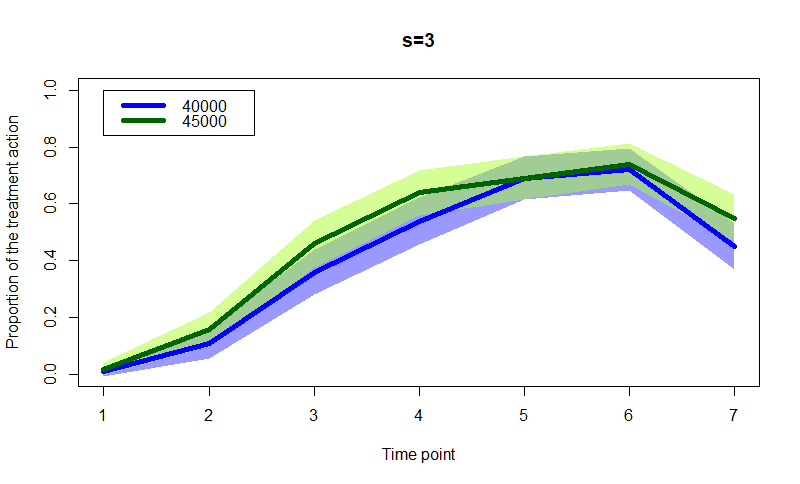}
    \caption{Proportion of treatment action vs observation time for patients belonging to income group Rs. 40000 and Rs. 45000 respectively}
    \label{40vs45}
\end{figure}

\begin{table}[ht]
    \centering
\begin{tabular}{c| c c c|c c c}
\hline\hline
\multirow{3}{*}{\begin{tabular}[c]{@{}c@{}}Time\\ ($N=8$)\end{tabular}} & \multicolumn{6}{c}{Cancer stages across income groups}                    
\\ \cline{2-7} & \multicolumn{3}{c|}{Rs. 40000} & \multicolumn{3}{c}{Rs. 45000} \\ \cline{2-7} 
 & stage 1  & stage 2  & stage 3  & stage 1  & stage 2  & stage 3  \\ \hline
1                                                                       & 0.01     & 0.02     & 0.01     & 0.03     & 0.03     & 0.02     \\ 
2                                                                       & 0.05     & 0.09     & 0.11     & 0.05     & 0.07     & 0.16     \\ 
3                                                                       & 0.08     & 0.18     & 0.36     & 0.07     & 0.15     & 0.46     \\ 
4                                                                       & 0.13     & 0.25     & 0.54     & 0.10     & 0.18     & 0.64     \\ 
5                                                                       & 0.16     & 0.30     & 0.69     & 0.10     & 0.20     & 0.69     \\ 
6                                                                       & 0.18     & 0.31     & 0.72     & 0.10     & 0.20     & 0.74     \\ 
7                                                                       & 0.09     & 0.17     & 0.45     & 0.08     & 0.15     & 0.55     \\ \hline\hline
\end{tabular}

    \caption{Proportion of action 2 (the treatment action) across different time points of treatment and stages of cancer for income groups Rs. 40000 and Rs. 45000 respectively}
    \label{table4}
\end{table}

We would like to observe how the financial status of a patient affects the optimal action suggested by our methodology. Higher is a patient's socio-economic stature, more he is able to afford the expensive treatment actions and consequently higher is his chance to recover from cancer. In this case the optimal action policies in that keep in mind a patient's health status as well as the socio-economic standing. We simulate 200 patients for two different income ($\theta$) groups, each  group with 100 patients who have same income and other covariate profile varying over their corresponding ranges. For each patient the optimal action matrix is calculated and subsequently we note what proportion of times action 2 (i.e. the treatment action) is declared as optimal.

It is clearly observable that greater the difference between the income levels is, more often the treatment action is suggested for the person with higher income. In the figure \ref{10vs80}, the dark green line indicates the proportion of action 2 over time for a person having income Rs. 80000. The green line is strictly above the blue line that denotes the evolution of action 2 over time for the a with
income Rs. 10000. Only for stage 3 cancer, the blue line intersects the green line at one point ($5 < t < 6$) and goes above the green line. Such behaviour can be interpreted as a final recovery effort near the end of the horizon shown by the person with lower income. That is, his objective near the end of the horizon should be to maximize his health related reward value (when the reward contributed by the income, as a result of its exponential decay property, is very low) even if the income is lower. This kind of phenomenon well matches with our intuition regarding a patient's health economic perspective of an expensive treatment modality. Moreover, on close observation of Table \ref{table1}, one may notice that there quite a few zeros in the table. Since the table plots the proportion of times the expensive treatment is suggested, this shows that when the expenses are high and socialized medicare is not prevalent, we are better off suggesting a cheaper treatment so as to keep the overall quality of life of the individual at the optimal level. 

Another interesting observation of our simulation study is that as the difference between the financial status goes down, the green and the blue lines start to approach towards each other. In other words, if the difference between the two income levels is lower, the difference in proportion of action 2 is also diminished. Furthermore, for each of the cases considered here, the proportion of action 2 is higher when a patient is suffering from stage 3 cancer. Therefore, our model is evidently susceptible towards a cancer patient's socio-economic stature as well as the severity of the cancer.

The \textbf{\texttt{R}} codes for the simulation of covariates, search for optimal treatment policies and plots are provided  
\href{https://github.com/navonil1729/MDP-and-QOL-in-Cancer-Treatment}{\texttt{https://github.com/navonil1729/MDP-and-QOL-in-Cancer-Treatment}}.

\section{Discussion}\label{conclusion}

Cancer is a major global hazard. Diagnosis of cancer is a field that is widely being explored by researchers involved in interdisciplinary fields varying over biological sciences, statistical sciences, computer science and many more. In this literature, we make gentle yet sincere attempt to point out how everyday life factors play great role in cancer treatment and we also formulate a statistical model that captures all those factors and tries to suggest the optimal treatment action. Our contribution in this paper is to reconcile the framework of dynamic treatment regimes and the emphasis on QOL factors. The factors like reliability and effectiveness of treatment, along with its expense and subsequent number of healthy days given, can be incorporated in the reward function of the Markov Decision Process we elucidated. Moreover, with an agreed upon choice of reward function, the remaining procedure is fully automated. Thus this framework produces an objective way to treat patients keeping in mind their entire QOL criterion.

In our simulation section, we use the financial status of patients as a proxy for affordability of medication in absence of insurance or medicare schemes. More wealthy individuals can generally afford to be more health conscious, and thus, have more chances of recovery. We have incorporated this in our simulation section to have a more realistic outcome as well. Based on the choice of the reward function, our optimal actions change in the subsequent outcome, which shows that our framework is indeed successful in incorporating these effects. Susceptibility to the exposure due to professional causes has been captured here in the indicator covariate of radioactive exposure. Similarly, mining and pottery experience, dust exposure, pollution levels can be incorporated into the training model for the transition probabilities, so they can be effectively estimated from data, and directly used for predicting medication. We have shown the proportion of times an optimal action profile of severe treatment is chosen, suitably marginalized on the other factors. This proportion has indeed been susceptible to income levels, in a sensitive and robust way. It's sensitive as the action profiles differ from one income level to other, but robust because as income levels come closer, the optimal action profiles turn out to be similar as well. Thus slight perturbations would not change the action profile dramatically. 

Further scopes of exploration in this regard include doing away with the Markovian set up. The Markov assumption is good if we divide the potential states a person can be as different strata, but that increases the dimensionality of the problem to being potentially intractable. An obvious way around that could be using a longer time dependence structure through more time periods. A useful generalization could be the use of Martingales, modeling the medical history a person accumulates as a sequence of filtration, and use suitable optimization techniques to obtain optimal outcomes. On the behavioral aspects, as more and more emphasis on QOL factors grow, the effect it has on the reward function become evident. Another issue is that of time homogeneity. We have assumed for simplicity that the stochastic process is time homogeneous, but the framework does not require that as a necessity. In fact, an easy way to incorporate time dependence structure in the probabilities is making the $\beta$-coefficients of the covariates  in the proportional odds cumulative logit model dependent on time as well. This can easily and efficiently eradicate any potential problem due to the assumption of time homogeneity. In fact, the model can be extended to a continuous time framework, where the arrival of the patient to the doctor is a random process, and the physician's decision are tuned according this probability. However, these approaches would require the physician to know the epoch of the disease process for the patient, as that would influence the physician's optimal course of actions. Such developments, in conjunction with behavioral techniques, can in future lead to suitable nudging techniques and behavioral therapy which can be evident on improving upon the reward function, and as a consequence, the overall quality of life of the patient. 

Based on our framework, the optimal treatment scheme is now dependent heavily on the choice of the reward function. The reward function must be optimally chosen, incorporating all the factors that can affect the overall well being of the individual. The physician's objective and the patient's objective must be amalgamated in formulating the reward function, with appropriate weightage on both the factors, so as to circumvent the problem of dropouts, incomprehensibility of instructions, and financial and socio-economic barriers to recovery.


\bibliographystyle{plain}
\bibliography{main.bbl}
\end{document}